\documentclass[summary]{ursi}

\usepackage{bm}
\usepackage{color}

\newcommand{\simnot}{\mathord{\sim}}

\title{A GPU-based Imager with Polarised Primary-beam Correction}

\author{Chris J. Skipper*\affref{ref1}, Anna M. M. Scaife\affref{ref1}\affref{ref2}, J. Patrick Leahy\affref{ref1}
  }

\affiliation{%
  \aff{ref1}{Jodrell Bank Centre for Astrophysics, The University of Manchester, Manchester, M13 9PL, UK}
  \aff{ref2}{The Alan Turing Institute, Euston Road, London, NW1 2DB, UK}
}

\begin{document}

\maketitle

\begin{abstract}
The next generation of radio telescopes will strive for unprecedented dynamic range across wide fields of view, and direction-dependent gains such as the gain from the primary-beam pattern, or leakage of one Stokes product into another, must be removed from the cleaned images if dynamic range is to reach its full potential. Unfortunately, such processing is extremely computationally intensive, and is made even more challenging by the very large volumes of data that these instruments will generate. Here we describe a new GPU-based imager, aimed primarily at use with the ASKAP telescope, that is capable of generating cleaned, full-polarisation images that include wide-field, primary-beam, and polarisation leakage corrections.
\end{abstract}

\section{Introduction}

Astronomical images that are made with interferometric radio telescopes have to be cleaned, or deconvolved, in order to remove the point-spread function of the interferometer. But other, weaker, effects that are direction-dependent, such as gains due to the antenna primary-beam patterns, are normally ignored due to the complexity of processing required to remove them. However, the latest - and next - generations of radio telescopes (such as the Square Kilometre Array (SKA), MeerKAT \cite{Jonas2009}, or ASKAP \cite{Johnston2007}) will need to clean out these direction-dependent gains if their full potential in dynamic range is to be reached.

Direction-dependent gains that are not frequency or time dependent can be removed in the image domain (providing these gains are known) with relatively little processing, but gains that vary with time or frequency have to be handled in the UV (or visibility) domain where different corrections can be applied to different visibilities. This processing is sometimes referred to as A-projection \cite{Bhatnagar2008,Mitchell2012}, and has been available in image-processing packages for some time, albeit not in wide use.

In this paper we describe a new GPU-based imager, aimed primarily at use with the ASKAP telescope, that is capable of generating cleaned, full-polarisation images that can correct for wide-field imaging effects (w-projection), frequency and direction-dependent primary-beam effects, and the leakage of flux from one polarisation product into another. The many thousands of cores available in modern GPUs can be fully utilised for tackling many aspects of radio imaging, making full-polarisation primary-beam gain and leakage correction a viable option for processing the enormous volumes of data that telescopes such as ASKAP or the SKA will be producing.

\section{Imaging capability}

\subsection{Overview}

The GPU imager described in this work reads data from a CASA\footnote{https://casa.nrao.edu/} \cite{McMullin2007} measurement set\footnote{Functionality to handle multi-measurement sets (.MMS) is not yet implemented, but will be made available in later releases.}, and generate a CASA image which can be viewed in the CASA viewer.  The output images are cleaned using an implementation of the Cotton-Schwab algorithm \cite{Schwab1984}, which accepts many of the usual parameters such as the number of clean cycles, the loop gain, the cycle factor, and a threshold stopping criterion.

Wide-field image correction is fully implemented using the w-projection algorithm \cite{Cornwell2005}, with a configurable number of w-planes. Primary-beam correction is also included (based upon the algorithm described in \cite{Bhatnagar2008}), and will correct for direction-dependent gains resulting from the primary-beam pattern, and leakage from one Stokes product into another.

\subsection{Mosaicing}

The GPU imager is able to construct mosaic images in either the UV or image domains, based upon multiple fields in a single measurement set, multiple measurement sets, or both of these at the same time.

For UV-domain mosaics, each mosaic component is weighted during gridding by using its primary beam as a gridding kernel, and these weighting factors are later removed from the image domain through multiplication of the image by
\begin{equation}
	\label{eqn-mosaic-correction}
	\gamma(l,m) = \Bigg[\frac{\sqrt{\sum^{N}_{i}{{B_{i}(l,m)}^2}}}{{\rm \max}\sqrt{\sum^{N}_{i}{{B_{i}(l,m)}^2}}}\Bigg]_{\textsc{pbp}} \Bigg[\frac{1}{\sum_i^N{{B_{i}(l,m)}^2}}\Bigg]_{\textsc{wc}},
\end{equation}
where $\textsc{pbp}$ is the primary-beam pattern normalised to a maximum of 1 over the image, $\textsc{wc}$ is the weighting correction for the mosaic components, $N$ is the number of mosaic components, and $B_{i}$ is the primary beam for mosaic component $i$.

For image-domain mosaics, an image of each mosaic component is generated with the phase position at its centre, and image-plane reprojection is used to construct the mosaic at a common phase position. Prior to the degridding stage of cleaning the model image is reprojected to the phase position of each mosaic component, and multiple FFT and degridding steps are performed. Since FFTs are very efficient on a GPU (more than two orders of magnitude faster than running the FFTW libraries on the CPU), the performance of image-domain mosaicing compares favourably with that of UV-domain mosaicing, which often requires much larger kernel sizes.

\subsection{Note on performance}

The performance of the GPU imager is clearly heavily dependent upon the hardware on which it is run, and therefore performance statistics can be misleading. However, in all tests conducted so far the GPU imager was found to be at least one order of magnitude faster than CASA \textsc{tclean}, with the greatest performance advantage (${\simnot 65\times}$ speedup over \textsc{tclean}) obtained from constructing mosaic images from data from the ALMA \cite{Wootten2009} telescope, in which the grid size is typically small and {w-projection} (for wide-field correction) is not usually required. Our tests have largely been conducted on a single Tesla K20 or K40 GPU, and compared to the performance of \textsc{tclean} on an Intel Core i5 or Intel Xeon E5 CPU (using the non-multithreaded version of \textsc{tclean}). New benchmarking tests currently being implemented on Nvidia A100 GPUs are expected to show further improvements in relative performance. 

\section{Primary-beam correction}

\subsection{Overview}

When high-dynamic-range images are required we can no longer afford to ignore effects such as the direction-dependent gain of the primary beam, and polarisation leakage from one Stokes product to another. Bhatnagar et al \cite{Bhatnagar2008} showed how such effects can be corrected for in the UV domain during deconvolution, using additional kernels based upon the beam patterns, and here we have implemented this algorithm in part to provide non-time-dependent primary-beam and leakage correction. Our algorithm is intended primarily for processing ASKAP data, for which the third axis on each antenna removes the time dependence of the direction-dependent effects, but it would also work well for a rotationally-symmetric beam pattern (e.g. Airy disk) on a receiver that lacks the third axis.

\subsection{The Mueller matrix and its inverse}

Following \cite{Bhatnagar2008}, we can represent the beam patterns and polarisation leakage in the visibility equation using the Mueller matrices \cite{Mueller1948}, ${{\rm M}_{ij,\nu}}$, such that
\begin{equation}
	\label{eqn-visibility-eqn}
	{\rm V}_{ij,\nu}^{\rm obs} = \int{{{\rm M}_{ij,\nu}}(\bm{\hat{s}}) {{\rm I}_{\nu}^{\rm sky}}(\bm{\hat{s}}) \exp\Big[ i2\pi \bm{\hat{s}} \cdot \bm{\hat{b}_{ij,\nu}} \Big]} d\bm{\hat{s}},
\end{equation}
where ${{\rm V}_{ij,\nu}}$ is the full-polarisation visibility 4\nobreakdash-vector, ${{\rm I}_{\nu}^{\rm sky}}$ is the full-polarisation sky image 4\nobreakdash-vector, ${\bm{\hat{b}_{ij,\nu}}}$ are the baseline vectors between antennas $i$ and $j$ (in units of wavelength), and $\nu$ is the frequency. The Mueller matrices are a set of ${4 \times 4}$ matrices, derived from the ${2 \times 2}$ antennae Jones matrices using ${{\rm M}_{ij,\nu} = {\rm J}_{i,\nu} \otimes {\rm J}_{j,\nu}}$, that contain the beam patterns for the polarisation products XX*, XY*, YX*, and YY* along the diagonals, and the leakage patterns between those products in the off-diagonal elements, such that the relationship between the true sky images and the observed images is given by
\begin{equation}
	\label{eqn-mueller-xx-xy-yx-yy}
	\begin{bmatrix}
		{I_{XX^{\ast}}} \\ {I_{XY^{\ast}}} \\ {I_{YX^{\ast}}} \\ {I_{YY^{\ast}}}
	\end{bmatrix}_{ij,\nu}^{\rm obs}
	=
	{\rm M}_{ij,\nu} \times
	\begin{bmatrix}
		I_{XX^{\ast}} \\ I_{XY^{\ast}} \\ I_{YX^{\ast}} \\ I_{YY^{\ast}}
	\end{bmatrix}_{\nu}^{\rm sky}.
\end{equation}

However, we would prefer to work with Stokes products I, Q, U and V rather than the polarisation products XX*, XY*, YX*, and YY*. Through a few summations and inverse operations, and the substitutions
\begin{equation}
	\label{eqn-stokes}
	\begin{split}
		{\rm Stokes\,I} &= \frac{XX^{\ast} + YY^{\ast}}{2}, \\
		{\rm Stokes\,Q} &= \frac{XX^{\ast} - YY^{\ast}}{2}, \\
		{\rm Stokes\,U} &= i \frac{XY^{\ast} + YX^{\ast}}{2}, \\
		{\rm Stokes\,V} &= i \frac{YX^{\ast} - XY^{\ast}}{2},
	\end{split}
\end{equation}
we can manipulate ${\rm M_{ij,\nu}}$ to give
\begin{equation}
	\label{eqn-mueller-stokes}
	\begin{bmatrix}
		I_{\rm\,I} \\ I_{\rm\,Q} \\ I_{\rm\,U} \\ I_{\rm\,V}
	\end{bmatrix}_{ij,\nu}^{\rm obs}
	=
	{\rm M}_{ij,\nu}^{\rm S} \times
	\begin{bmatrix}
		I_{\rm\,I} \\ I_{\rm\,Q} \\ I_{\rm\,U} \\ I_{\rm\,V}
	\end{bmatrix}_{\nu}^{\rm sky},
\end{equation}
and it is this form for the Mueller matrices that provides the gridding convolution kernels required for transforming a model image into a set of visibilities during the `degridding' stage of Cotton-Schwab clean (Eqn.~\ref{eqn-visibility-eqn}). The degridding kernels are generated by the fast Fourier transform of the Mueller matrix cells from the image domain into the UV domain, and typically are first multiplied in the image domain by an anti-aliasing function and the w\nobreakdash-correction term.

For the `gridding' stage, in which the true sky images are recovered from a set of visibilities, we need to compute the inverse of the Mueller matrices, ${{{\rm M}_{ij,\nu}^{\rm S}}^{-1}}$, which (once transformed to the UV domain) provide the gridding convolution kernels required to recover the true sky image (in Stokes I, Q, U, or V) from the observed visibilities, removing the primary beam and leakage patterns.

Note that although the Mueller matrices are functions of $i$, $j$, and $\nu$, and are therefore unique to each baseline and channel, for performance reasons it is likely that the beam patterns will need to be averaged over a range of channels, and potentially over multiple baselines as well.

\subsection{Handling the nulls in the primary beam}

Since the intensity of the primary beam pattern approaches zero at the nulls, it follows that the inverse, which provides the gridding kernels, will get very large at those positions. Although the final image will often be masked where the primary beam is weak, these large kernel values can still cause a number of problems during the gridding and FFT stages. For instance, the highest-frequency channels have a primary beam that may be narrower than the one used to apply the image mask, and the gridding kernel computed from its inverse can result in the appearance of a `noisy ring' around the edges of the image.

We therefore apply a correcting function to the primary beam patterns that prevents the gridding kernels from becoming too large. Working on copies of the Mueller matrices (not the originals, which are required for computing the degridding kernels), the primary beam patterns ${B_{\rm I}}$, ${B_{\rm Q}}$, ${B_{\rm U}}$ and ${B_{\rm V}}$ from the diagonal elements are modified using
\begin{equation}
	B_p^\prime(l,m) = \frac{B_p(l,m)}{\lvert B_p(l,m) \rvert} \Bigg[ \lvert B_p(l,m) \rvert + \frac{\beta}{\lvert B_p(l,m) \rvert} \Bigg],
	\label{eqn-modified-mueller}
\end{equation}
where $\beta$ is given by ${((\epsilon / 2) \times \max{\lvert B_p \rvert})^2}$, and $\epsilon$ is fraction of the beam intensity below which the image is masked (typically 0.2). We then calculate the inverse Mueller matrix, ${{{\rm M}_{ij,\nu}^{\rm S}}^{-1}}$, from which we can extract the gridding convolution kernels.

\begin{figure}
  \centering
  \includegraphics[width=80mm]{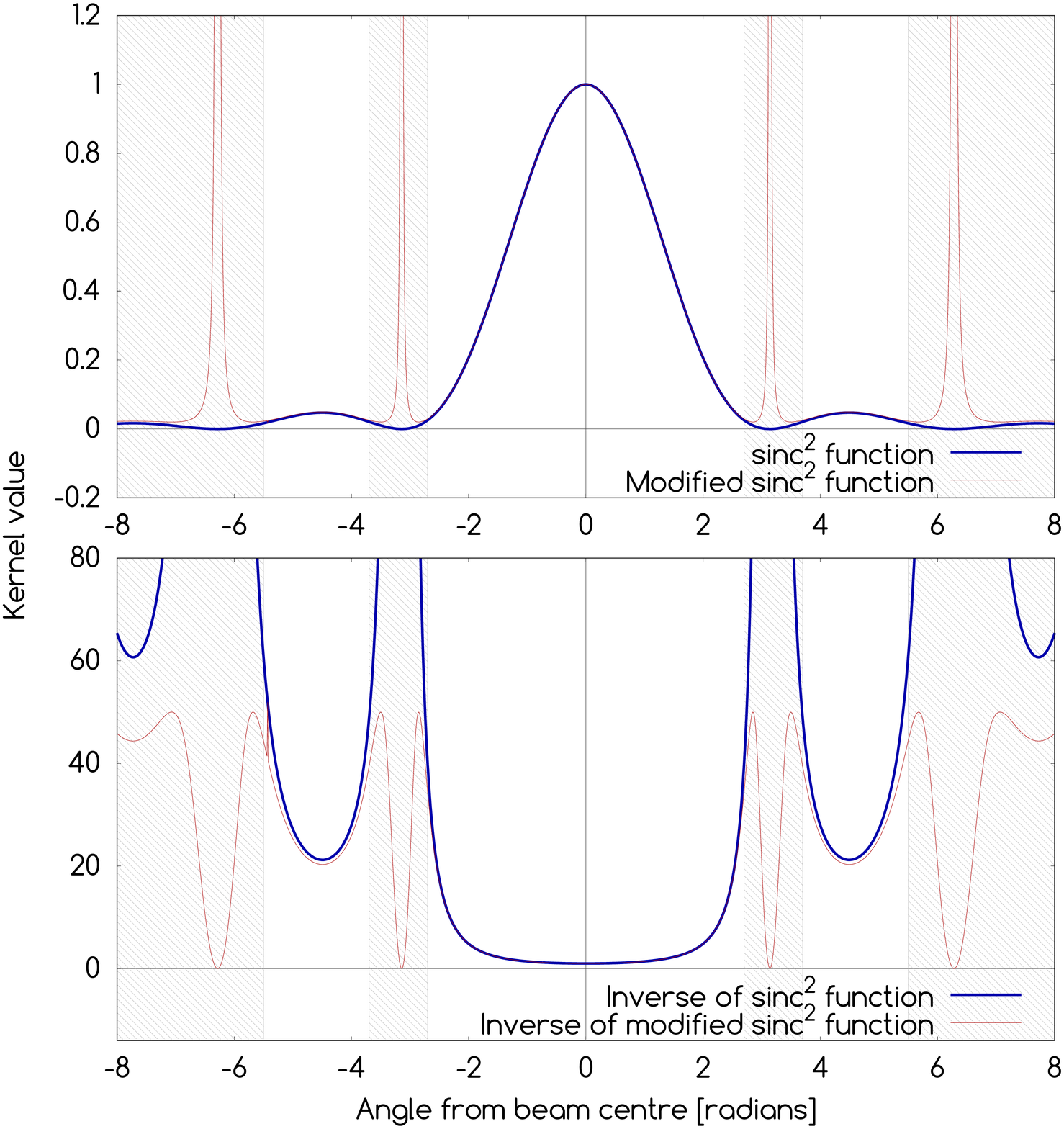}
  \caption{Top: A primary beam, modelled with a ${sinc^2}$ function, is modified by Eqn.~\ref{eqn-modified-mueller} so that the function does not approach zero at the null points of the beam. The shaded areas show regions of the telescope field of view that will be masked out of the image due to the beam amplitude being less than 2~per~cent. Bottom: The inverse of the modified primary beam, which is used in gridding to correct for the shape of the beam, is not afflicted with the discontinuities and infinities that can be seen in the inverse of the unmodified primary beam, yet both functions track each other closely in the unmasked area of the image.}
  \label{fig-sinc-function}
\end{figure}

Fig.~\ref{fig-sinc-function} (top panel) shows how modifying a primary beam pattern from the Mueller matrices so that it does not approach zero at the null points of the beam ensures that its inverse (bottom panel) is suitable for use as a gridding convolution kernel. The shaded areas in Fig.~\ref{fig-sinc-function} show regions of the telescope field of view where the beam amplitude is less than 2~per~cent, and the images produced will be masked (this fraction is more typically 20~per~cent, but here we set a lower value in order to include the first sidelobes of the ${\rm sinc^2}$ function); in the unshaded region we can see that the modified beam pattern closely matches the unmodified beam pattern, but is not afflicted with the discontinuities and infinities that affect the latter in the shaded regions. At the centre of the beam the inverse of the modified function diverges from the inverse of the unmodified function by a factor of ${10^{-4}}$; at the half-beam-power angle this divergence is ${4 \times 10^{-3}}$, and at the first sidelobe ${4 \times 10^{-2}}$.

\section{Conclusion}

The new GPU-based imager described in this paper has shown encouraging results so far, both in terms of image quality and speed of execution. There are many aspects of radio imaging that are ideally suited to processing on a highly parallised architecture, including FFTs, gridding, kernel generation, image-plane reprojection, and phase rotation; conversely, more iterative processes such as H{\"o}gbom cleaning \cite{Hogbom1974} are less able to take full advantage of parallelisation.

Although there are other options for correcting direction-dependent effects in ASKAP, MeerKAT, or SKA data, including faster but less accurate image-domain corrections, it is clear that GPUs can offer the processing power that allows primary-beam and leakage correction to be performed in non-prohibitive time frames. However, given that existing radio-astronomy software, including that for calibration, flagging, etc, is not GPU-based then much would depend upon how a GPU-based imager could be effectively integrated into new and existing data-reduction pipelines. 

\section{Acknowledgements}

CS gratefully acknowledges support from the UK Research \& Innovation (UKRI) Science \& Technology Facilities Council (STFC) under grant ST/T000414/1. AMS gratefully acknowledges support from Alan Turing Institute AI Fellowship EP/V030302/1. 


\end{document}